\documentclass[a4paper,fleqn]{cas-sc}
\usepackage{makecell}
\usepackage{amsmath}
\usepackage{bm}
\usepackage{pifont}
\usepackage{algorithm}  
\usepackage{algorithmic}
\usepackage{threeparttable}
\usepackage{color}
\usepackage{appendix}
\usepackage{listings}

\usepackage[numbers, compress]{natbib}
\def\tsc#1{\csdef{#1}{\textsc{\lowercase{#1}}\xspace}}
\tsc{WGM}
\tsc{QE}
\begin{document}
\let\WriteBookmarks\relax
\def\floatpagepagefraction{1}
\def\textpagefraction{.001}

% Short title
\shorttitle{A modified Newmark/Newton-Raphson method with automatic differentiation for general nonlinear dynamics analysis}

% Short author
\shortauthors{Yifan Jiang et~al.}

% Main title of the paper
\title [mode = title]{A modified Newmark/Newton-Raphson method with automatic differentiation for general nonlinear dynamics analysis}

% Title footnote mark
% eg: \tnotemark[1]
\tnotemark[1]

% Title footnote 1.
% eg: \tnotetext[1]{Title footnote text}

% First author
%
% Options: Use if required
% eg: \author[1,3]{Author Name}[type=editor,
%       style=chinese,
%       auid=000,
%       bioid=1,
%       prefix=Sir,
%       orcid=0000-0000-0000-0000,
%       facebook=<facebook id>,
%       twitter=<twitter id>,
%       linkedin=<linkedin id>,
%       gplus=<gplus id>]

\author[1]{Yifan Jiang}[type=editor,
      style=chinese
]
\fnmark[1]

\author[1]{Yuhong Jin}[type=editor,
      style=chinese
]
\fnmark[1]
% Corresponding author indication
% \cormark[1]

% Footnote of the first author
% \fnmark[1]

% Email id of the first author
% \ead{<email address>}

% URL of the first author
% \ead[url]{<URL>}

% Credit authorship
% eg: \credit{Conceptualization of this study, Methodology, Software}
% \credit{<Credit authorship details>}

% Address/affiliation
% \affiliation[<aff no>]{organization={},
%             addressline={}, 
%             city={},
% %          citysep={}, % Uncomment if no comma needed between city and postcode
%             postcode={}, 
%             state={},
%             country={}}

\author[1]{Lei Hou}[type=editor,
      style=chinese,
      orcid=0000-0003-0271-7323
]
\cormark[1]

% Footnote of the second author
% \fnmark[2]

% Email id of the second author
\ead{houlei@hit.edu.cn}

% URL of the second author
\ead[url]{http://homepage.hit.edu.cn/houlei}

\author[2]{Yi Chen}[type=editor,
      style=chinese
]

\author[1]{Andong Cong}[type=editor,
      style=chinese
]

% \author[1]{Haiming Yi}[type=editor,
%       style=chinese
% ]

% \author[1]{Zhonggang Li}[type=editor,
%       style=chinese
% ]
% \cormark[1]
% \ead{lizhonggang2001@163.com}

% \author[3]{Yongzhi Feng}[type=editor,
%       style=chinese
% ]

% \author[4]{Shun Zhong}[type=editor,
%       style=chinese
% ]

% Credit authorship
% \credit{1111}

% Address/affiliation
% \affiliation[1]{organization={11},
%             addressline={11}, 
%             city={11},
% %          citysep={}, % Uncomment if no comma needed between city and postcode
%             postcode={111}, 
%             state={11},
%             country={11}}

% Corresponding author text
\cortext[1]{Corresponding author}

% Footnote text
\fntext[1]{These authors contributed equally to this work.}

% Address/affiliation
\affiliation[1]{organization={School of Astronautics},
      addressline={Harbin Institute of Technology},
      city={Harbin},
      %          citysep={}, % Uncomment if no comma needed between city and postcode
      postcode={150001},
      % state={11},
      country={P. R. China}}
\affiliation[2]{organization={School of Civil Engineering},
      addressline={Harbin Institute of Technology},
      city={Harbin},
      %          citysep={}, % Uncomment if no comma needed between city and postcode
      postcode={150090},
      % state={11},
      country={P. R. China}}
% \affiliation[2]{organization={AECC Shengyang Engine Research Institute},
%       city={Shenyang},
%       %          citysep={}, % Uncomment if no comma needed between city and postcode
%       postcode={110000},
%       % state={11},
%       country={P. R. China}}
% \affiliation[3]{organization={Harbin Electric Company Limited},
%       city={Harbin},
%       %          citysep={}, % Uncomment if no comma needed between city and postcode
%       postcode={150001},
%       % state={11},
%       country={P. R. China}}
% \affiliation[4]{organization={Department of Mechanics},
%       addressline={Tianjin University},
%       city={Tianjin},
%       %          citysep={}, % Uncomment if no comma needed between city and postcode
%       postcode={300072},
%       % state={11},
%       country={P. R. China}}

% For a title note without a number/mark
%\nonumnote{}
% Here goes the abstract
\begin{abstract}
      The Newmark/Newton-Raphson (NNR) method is widely employed for solving nonlinear dynamic systems. However, the current NNR method exhibits limited applicability in complex nonlinear dynamic systems, as the acquisition of the Jacobian matrix required for Newton iterations incurs substantial computational costs and may even prove intractable in certain cases. To address these limitations, we integrate automatic differentiation (AD) into the NNR method, proposing a modified NNR method with AD (NNR-AD) to significantly improve its capability for effectively handling complex nonlinear systems. We have demonstrated that the NNR-AD method can directly solve dynamic systems with complex nonlinear characteristics, and its accuracy and generality have been rigorously validated. Furthermore, automatic differentiation significantly simplifies the computation of Jacobian matrices for such complex nonlinear dynamic systems. This improvement endows the NNR method with enhanced modularity, thereby enabling convenient and effective solutions for complex nonlinear dynamic systems.
\end{abstract}

% Use if graphical abstract is present
%\begin{graphicalabstract}
%\includegraphics{}
%\end{graphicalabstract}

% Research highlights
\begin{highlights}
      \item The Newmark/Newton-Raphson method is improved through the integration of automatic differentiation.
      \item The modified method demonstrates advantages of simplicity, high efficiency, and modularity.
      \item The modified method exhibits both accuracy and universality in solving complex nonlinear dynamic systems.
      \item We provide a general open-source toolbox for solving complex nonlinear dynamic systems.
\end{highlights}

% Keywords
% Each keyword is seperated by \sep
\begin{keywords}
      \sep Automatic differentiation \sep Newmark/Newton-Raphson \sep Nonlinear dynamics \sep Ordinary differential equation
\end{keywords}

\maketitle

% Main text
\section{Introduction}

The mathematical formulation of a system's dynamic model is typically represented by second-order ordinary differential equations (ODEs). Solving these equations to obtain the system's response is of significant importance for understanding and analyzing its dynamic characteristics. In 1959, Newmark proposed the Newmark method for solving dynamics problems by establishing approximate relationships among displacement, velocity, and acceleration \cite{newmark_method_1959}. The various numerical characteristics and parameter selections of the Newmark method \cite{fox_new_1949, hughestjr_precis_1983}, along with its local and global errors \cite{brown_discretization_1985,warburton_assessment_1990}, have been extensively studied by scholars. 

Based on the Newmark method, numerous scholars have proposed various algorithmic improvements tailored to the specific analytical requirements of different fields. Fan proposed an iterative algorithm for dynamic load identification based on the Newmark method, which, compared to the explicit Newmark method, demonstrates higher  recognition precision and calculation efficiency \cite{yuchuan_dynamic_2020}. Benítez introduced non-Eulerian Newmark method for solving the mechanical response of free-boundary continua \cite{benitez_noneulerian_2020}. Sheng proposed a stability-improved spectral-element time-domain method based on the Newmark method, which has a significant effect on the suppression of late-time instability \cite{sheng_stabilityimproved_2019}. Pasetto introduced a waveform relaxation Newmark method for solving linear second-order hyperbolic ordinary differential equations, which retains the unconditional stability of the implicit Newmark method with the advantage of the lower computational cost of explicit time integration schemes \cite{pasetto_waveform_2019}. Zhang developed a new scheme based on the Newmark method by expanding the displacement to the fourth derivative, thereby formulating a four-stage direct integration scheme for structural dynamics analysis \cite{zhang_fourstage_2017}. Miao proposed a sub-cycling technique based on the Newmark method, which is well-suited for solving condensed flexible multi-body dynamics models \cite{miao_study_2008}. Bui introduced a modified Newmark method for nonlinear dynamic models, which enhances unconditional stability and ensures dissipation in the nonlinear regime \cite{bui_modified_2004}. Zhang proposed and developed a stochastic Newmark method for the random analysis of multi-degree-of-freedom nonlinear systems, based on the deterministic Newmark integration formula \cite{zhang_stochastic_1999}. 

Due to the demonstrated stability and versatility of the Newmark method and its modified versions in solving ODEs, numerous scholars have successfully employed these approaches to analyze a wide range of engineering problems. Sun employed a combined approach utilizing the Newmark method and the Heaviside step function to calculate the dynamic response of a beam incorporating a nonlinear energy sink \cite{sun_vibration_2025}. Li utilized the Newmark method to solve the dynamic response of an inverted Stewart platform with a flexible base \cite{li_rigidflexible_2025}. Sudip employed the Newmark method to evaluate the vibration reduction capability of the nonlinear stiffened inertial amplifier tuned mass friction damper \cite{chowdhury_nonlinear_2025}. Liu conducted a nonlinear dynamic analysis of three-dimensional geometrically exact curved beams using the Newmark-Wilson method \cite{liu_nonlinear_2024}. Fu employed the dynamic Newmark method to calculate the dynamic safety factor and permanent displacement of rock blocks under seismic action \cite{fu_extensions_2021}. Miles utilized the Newmark method to investigate rigorous landslide hazard zonation \cite{miles_rigorous_1999}.

The conventional Newmark method is exclusively applicable to linear systems and becomes ineffective when dealing with nonlinear systems. To overcome this limitation, Jacob proposed the Newmark/Newton-Raphson method and elaborated its comprehensive computational procedures along with implementation details \cite{jacob_optimized_1994}. Ma utilized this method to analyze the vibration response of a rotor system considering coupling misalignment and nonlinear contact between the shaft and disk \cite{ma_vibration_2025}. Zanussi employed this method to investigate the nonlinear flutter of functionally graded carbon nanotube-reinforced composite quadrilateral plates \cite{pashazanussi_nonlinear_2024}. Saimi used this method to obtain the excitation force response of a symmetrical on-board rotor mounted on hydrodynamic bearings \cite{saimi_nonlinear_2023}. Miao employed this method to analyze the nonlinear vibrations of a variable-speed rotor system under maneuvering flight conditions \cite{miao_nonlinear_2023}. Wei utilized this method to capture the transient mechanical response of simply supported sandwich plates subjected to dynamic blast \cite{wei_large_2022}. Huang applied this method to obtain the dynamic response of the shaft-bearing system of a marine propeller shaft \cite{huang_dynamical_2019}. Zhong investigated the nonlinear dynamic response of functionally graded beams on a tensionless elastic foundation under thermal shock based on this method \cite{zhong_analysis_2016}. Liu utilized this method to study the galloping of iced conductors \cite{liu_nonlinear_2009}. Rubin employed this method to investigate finite rotations \cite{rubin_simplified_2007}. Harsha used this method to study the dynamic response of rolling bearings \cite{harsha_nonlinear_2004} and the dynamic response \cite{harsha_nonlinear_2005, harsha_nonlinear_2004a} and stability \cite{harsha_stability_2004} of rotor systems with rolling bearing supports. Hussain employed this method to investigate the dynamic response of two rotors connected by a rigid mechanical coupling \cite{al-hussain_dynamic_2002}. Meek utilized this method to study the dynamic response of beams considering geometric nonlinearity \cite{meek_nonlinear_1995}. Chen adopted this method to calculate the nonlinear response of graphene platelets-reinforced metal foam piezoelectric beams under various loading conditions \cite{chen_dynamic_2025}.

For high-degree-of-freedom systems with complex nonlinearities, obtaining the Jacobian matrix required for Newton iterations in the Newmark/Newton-Raphson method presents significant challenges. Specifically: (1) manual differentiation is computationally tedious and error-prone; (2) symbolic differentiation suffers from expression swell issues; and (3) numerical differentiation introduces roundoff errors and truncation errors. With the advancement of machine learning and the development of JAX and PyTorch \cite{ansel_pytorch_2024}, automatic differentiation (AD) has emerged as an increasingly favored alternative approach for derivative computation. AD is a technique that can automatically compute the derivatives of a computer program. Nolan first proposed the idea of using computer programs for AD in his doctoral thesis in 1953 \cite{nolan_analytical_1953}, which is the core concept of AD. Based on this concept, two distinct approaches to AD have emerged: forward mode \cite{wengert_simple_1964} and reverse mode \cite{linnainmaa_taylor_1976}. AD is composed of a finite set of basic operations, and the derivatives of these operations are known. By applying the chain rule to combine the derivatives of the constituent operations, the derivative of the entire composition can be obtained. Consequently, AD effectively overcomes the limitations inherent in the three aforementioned differentiation approaches, offering a robust solution to their respective shortcomings \cite{baydin_automatic_2018}. Due to these advantages, AD is now widely used in machine learning and has found extensive applications in various engineering fields such as solid mechanics \cite{pundir_simplifying_2025,rothe_automatic_2015}, heat transfer problems \cite{niu_automatic_2021,cai_physicsinformed_2021}, quantum physics \cite{guo_scheme_2021}, multibody dynamics \cite{callejo_direct_2019,callejo_performance_2014,callejo_hybrid_2014}, optimization \cite{norgaard_applications_2017,iri_roles_1997}, rotor dynamics \cite{furst_application_2016}, power system \cite{wei_wind_2013,ibsais_role_1997}, computational fluid dynamics \cite{jones_preparation_2011,muller_performance_2005}.

In this paper, we introduce AD into the Jacobian matrix computation of Newton-Raphson iterations, thereby proposing a modified Newmark/Newton-Raphson method with AD that effectively overcomes the drawbacks of manual differentiation (complexity, proneness to errors, and occasional intractability), circumvents the limitations of symbolic differentiation ( expression swell issues), and eliminates truncation and roundoff errors inherent in numerical differentiation, significantly enhancing the modularity and implementation feasibility of the Newmark/Newton-Raphson method for complex nonlinear dynamic systems.

The rest paper is organized as follows. Section \ref{sec:met} details the computational workflow and implementation specifics of AD and the modified Newmark/Newton-Raphson method with AD. Section \ref{sec:num} validates the versatility and accuracy of the proposed method through numerical examples: simple typical nonlinear systems, a system with complex nonlinear stiffness, a system with complex nonlinear stiffness and damping, and a high-dimensional system with nonlinear stiffness. Section \ref{sec:dic} presents an extended discussion of the proposed method. Finally, Section \ref{sec:con} concludes the paper and provides remarks.

\section{Methods}
\label{sec:met}

\subsection{Automatic differentiation}

The function $f:\mathbb{R}{^n}\to\mathbb{R}{^m}$ is defined as mapping an \textit{n}-dimensional input \textit{x} to an \textit{m}-dimensional output. The Jacobian matrix ${J_{ij}} = \partial {f_i}/\partial {x_j}$ , where $i =(1,...,m)$ and $j =(1,...,n)$, represents the derivative of the output with respect to the input. Automatic differentiation decomposes a complex function \textit{f} into elementary functions with known exact derivatives and computes the precise derivative of \textit{f} through basic arithmetic operations such as addition and subtraction along with the chain rule of calculus. For example, if the function \textit{f} can be decomposed as $f = h(k(x))$, such that $x \in \mathbb{R}{^n}$, $k:\mathbb{R}{^n}\to\mathbb{R}{^l}$and $h:\mathbb{R}{^l}\to\mathbb{R}{^m}$, the Jacobian matrix can be computed as 
\begin{equation}
  {J_{ij}} = \frac{{\partial {f_i}}}{{\partial {x_j}}} = \sum\limits_{g = 1}^l {\frac{{\partial {h_i}}}{{\partial {k_g}}}} \frac{{\partial {k_g}}}{{\partial {x_j}}}.
\end{equation}

\begin{figure}[pos=htbp]
	\centering
	\includegraphics[width=1.0\textwidth]{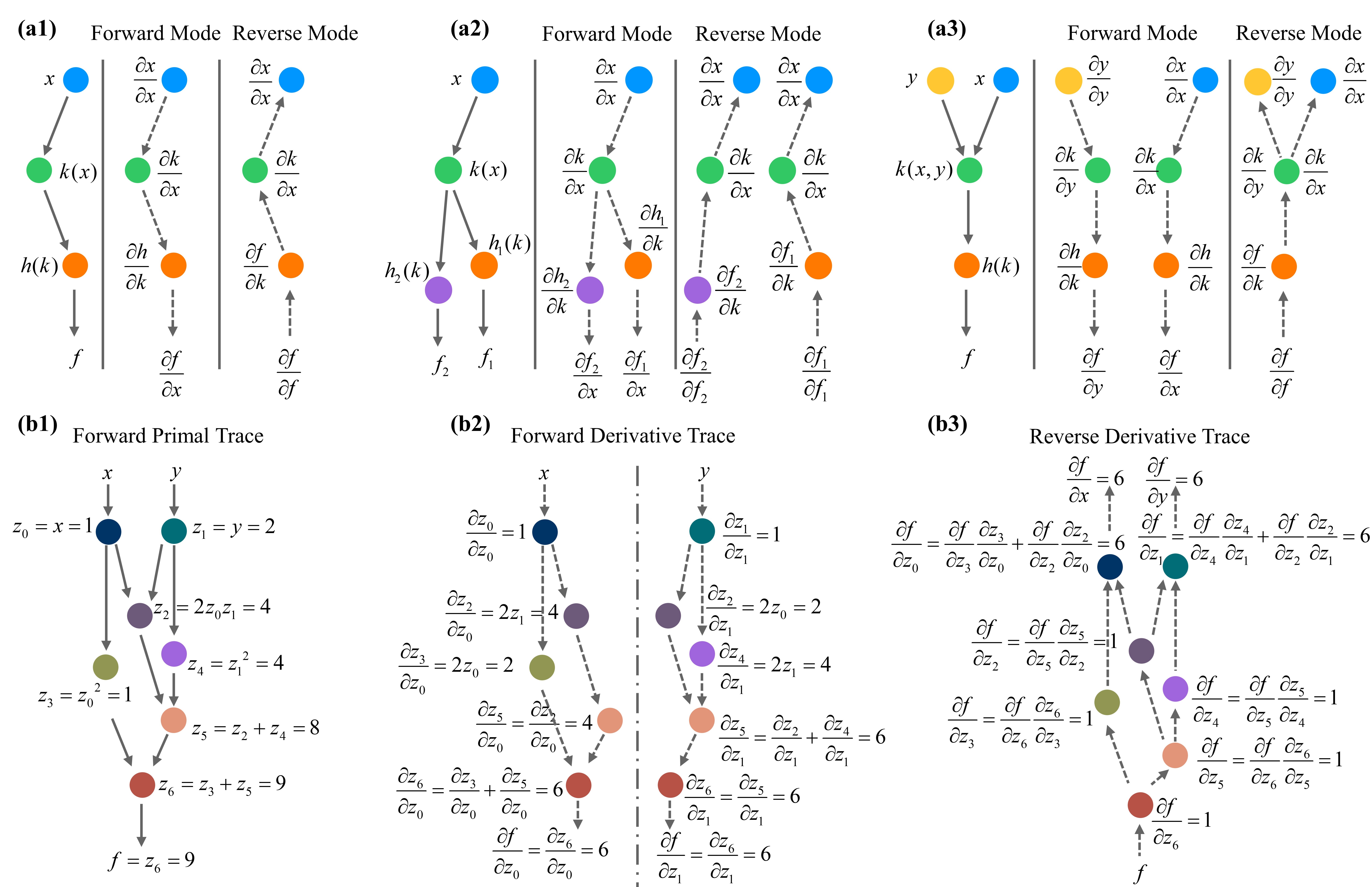}
	\caption{The computational graph of the function and its automatic differentiation. (\textbf{a1}) Function $f:\mathbb{R}\to\mathbb{R}$. (\textbf{a2}) Function $f:\mathbb{R}{^1}\to\mathbb{R}{^2}$. (\textbf{a3}) Function $f:\mathbb{R}{^2}\to\mathbb{R}{^1}$. (\textbf{b1}) The forward primal trace of function $f=x{^2}+2xy+y{^2}$. (\textbf{b2}) The forward derivative trace of function $f=x{^2}+2xy+y{^2}$. (\textbf{b3}) The reverse derivative trace of function $f=x{^2}+2xy+y{^2}$.}
	\label{fig:ad}
\end{figure}

Since automatic differentiation determines the derivative of \textit{f} by utilizing the exact derivatives of its subfunctions, it is unaffected by issues such as numerical instability or rounding errors. Even when the function \textit{f} is highly nonlinear or non-monotonic, automatic differentiation can compute derivatives with full precision, maintaining the same accuracy as theoretically derived analytical results.The computational graphs for forward mode and reverse mode automatic differentiation are illustrated in Fig.\ref{fig:ad}(a1). Forward mode automatic differentiation computes derivatives simultaneously with the forward propagation through the computational graph. Reverse mode automatic differentiation requires a forward pass through the computational graph to compute output values, followed by a backward propagation pass to calculate derivatives. The choice between these modes depends on the relative dimensions of the input and output spaces. As illustrated in Fig.\ref{fig:ad}(a2), when \textit{n} = 1 and \textit{m} = 2 (where \textit{n} < \textit{m}), the forward mode can compute both derivatives $\partial {f_1}/\partial {x}$ and $\partial {f_2}/\partial {x}$ in a single calculation, whereas the reverse mode requires two backward passes to obtain these derivatives. As shown in  Fig.\ref{fig:ad}(a3), when \textit{n} = 2 and \textit{m} = 1 (where \textit{n} > \textit{m}), the forward mode requires two separate computations to obtain both derivatives $\partial {f}/\partial {x}$ and $\partial {f}/\partial {y}$, while the reverse mode achieves this with just a single backward computation  \cite{pundir_simplifying_2025}. Therefore, forward mode differentiation is preferable when the output dimension exceeds the input dimension (\textit{m}> \textit{n}), whereas reverse mode differentiation becomes more efficient when the output dimension is significantly smaller than the input dimension (\textit{m} << \textit{n}). Considering the case in Fig.\ref{fig:ad}(a3), let $k(x,y)=x+y$ and $h(k)=k^2$, i.e., $f=h(k(x,y))=(x+y)^2=x^2+2xy+y^2$. The forward primal trace of $f(x,y)$ is then given in Fig.\ref{fig:ad}(b1). According to the computational method of automatic differentiation, $f$ is decomposed into a combination of simpler functions $z_i$. Through the forward mode of automatic differentiation (see Fig.\ref{fig:ad}(b2)), $\partial {f}/\partial {x}$ and $\partial {f}/\partial {y}$ can be obtained after two computations. Alternatively, via the reverse mode of automatic differentiation (see Fig.\ref{fig:ad}(b3)), $\partial {f}/\partial {x}$ and $\partial {f}/\partial {y}$ can be derived in a single computation.

\subsection{Newmark/Newton-Raphson method}

For ease of understanding and manuscript integrity, this subsection briefly introduces the computational workflow of the classical Newmark/Newton-Raphson (NNR) method \cite{jacob_optimized_1994}. Without loss of generality, any nonlinear dynamic equation can be expressed in the following form:
\begin{equation}
      \textbf{M}\ddot{\textbf{x}} + \textbf{C}\dot{\textbf{x}} + \textbf{Kx} + \textbf{F}(\ddot{\textbf{x}},\dot{\textbf{x}},\textbf{x}) = \textbf{Q}(t),
      \label{equ:equ1}
\end{equation}
where, $\textbf{M}$, $\textbf{C}$, and $\textbf{K}$ represent the linear mass matrix, damping matrix, and stiffness matrix, respectively. $\textbf{x}$ is the solution to the equation, representing the dynamic response of the system. $\textbf{F}(\ddot{\textbf{x}},\dot{\textbf{x}},\textbf{x})$ denotes the nonlinear forces within the system, including nonlinear mass effects $\textbf{F}(\ddot{\textbf{x}})$, nonlinear damping $\textbf{F}(\dot{\textbf{x}})$, and nonlinear stiffness $\textbf{F}({\textbf{x}})$. $\textbf{Q}(t)$ represents the external excitation applied to the system.

Assuming the system responses at time step $t_n$ $(\ddot{\textbf{x}}_n,\dot{\textbf{x}}_n,\textbf{x}_n)$ are known, the following equality holds according to Eq. (\ref{equ:equ1}):
\begin{equation}
      \textbf{M}\ddot{\textbf{x}}_n + \textbf{C}\dot{\textbf{x}}_n + \textbf{Kx}_n + \textbf{F}(\ddot{\textbf{x}}_n,\dot{\textbf{x}}_n,\textbf{x}_n) = \textbf{Q}(t_n).
      \label{equ:equ2}
\end{equation}

For the system responses at the next time step $t_{n+1}$ $(\ddot{\textbf{x}}_{n+1}, \dot{\textbf{x}}_{n+1}, \textbf{x}_{n+1})$, the following must also satisfy Eq. (\ref{equ:equ1}):
\begin{equation}
      \textbf{M}\ddot{\textbf{x}}_{n+1} + \textbf{C}\dot{\textbf{x}}_{n+1} + \textbf{Kx}_{n+1} + \textbf{F}(\ddot{\textbf{x}}_{n+1},\dot{\textbf{x}}_{n+1},\textbf{x}_{n+1}) = \textbf{Q}(t_{n+1}),
      \label{equ:equ3}
\end{equation}
moreover, we define the time interval between two adjacent steps as $\varDelta t$.

The Newmark method introduces the following two fundamental assumptions for the system responses between adjacent time steps ($t_n$ and $t_{n+1}$):
\begin{equation}
      \ddot{\textbf{x}}_{n+1} = \frac{1}{\beta \varDelta t^2} (\textbf{x}_{n+1}-\textbf{x}_n) - \frac{1}{\beta \varDelta t} \dot{\textbf{x}}_{n} - (\frac{1}{2\beta}-1) \ddot{\textbf{x}}_{n} ,
      \label{equ:equ4}
\end{equation}

\begin{equation}
      \dot{\textbf{x}}_{n+1} = \frac{\delta }{\beta \varDelta t} (\textbf{x}_{n+1}-\textbf{x}_n) +(1- \frac{\delta }{\beta}) \dot{\textbf{x}}_{n} + (1-\frac{\delta }{2\beta}) \varDelta t\ddot{\textbf{x}}_{n} ,
      \label{equ:equ5}
\end{equation}
where, $\beta$ and $\delta $ are parameters given by the Newmark method, $\beta$ determines stability and accuracy and $\delta $ controls numerical damping.

Substituting Eq. (\ref{equ:equ4}) and Eq. (\ref{equ:equ5}) into Eq. (\ref{equ:equ3}) yields the following algebraic equation:
\begin{equation}
      \textbf{R}(\textbf{x}_{n+1},\textbf{x}_n,\dot{\textbf{x}}_{n},\ddot{\textbf{x}}_{n},t_{n+1})=0 .
      \label{equ:equ6}
\end{equation}

For Eq. (\ref{equ:equ6}), $\textbf{x}_{n+1}$ is the only unknown variable, while all other quantities $(\textbf{x}_n,\dot{\textbf{x}}_{n},\ddot{\textbf{x}}_{n},t_{n+1})$are known. Therefore, the Newton-Raphson iteration method is employed to solve for $\textbf{x}_{n+1}$. First, we specify an initial value $\textbf{x}^0_{n+1}(k=0)$. Then, the Newton-Raphson iteration $\varDelta \textbf{x}\leftarrow \textbf{J}^{-1}\textbf{R}(\textbf{x}^k_{n+1},\textbf{x}_n,\dot{\textbf{x}}_{n},\ddot{\textbf{x}}_{n},t_{n+1})$ is performed to obtain $\varDelta \textbf{x}$, and the value of $\textbf{x}^k_{n+1}$ is updated via $\textbf{x}^k_{n+1}\leftarrow \textbf{x}^k_{n+1}-\varDelta \textbf{x}$, where \textbf{J} is the Jacobian matrix, defined as follows:
\begin{equation}
\textbf{J}=\frac{{\partial {\textbf{R}}}}{{\partial {\textbf{x}_{n+1}}}} | _{\textbf{x}_{n+1}=\textbf{x}^k_{n+1}}.
      \label{equ:equ7}
\end{equation}

Then, the Newton-Raphson iteration process is repeated until a convergence criterion is met, such as the Frobenius norm of the residual vector $\textbf{R}$ or increment $\varDelta \textbf{x}$ being smaller than a predefined threshold $\epsilon$. The final solution $\textbf{x}_{n+1}$ can then be output, and $\ddot{\textbf{x}}_{n+1}$ and $\dot{\textbf{x}}_{n+1}$ can be obtained through Eq. (\ref{equ:equ4}) and Eq. (\ref{equ:equ5}). Then, starting from the system response $(\ddot{\textbf{x}}_{n+1}, \dot{\textbf{x}}_{n+1}, \textbf{x}_{n+1})$ at time $t_{n+1}$, the aforementioned procedure is repeated to compute the system responses at all subsequent time steps.

\subsection{Modified Newmark/Newton-Raphson method with AD and its implementation}

\begin{figure}[pos=H]
	\centering
	\includegraphics[width=0.95\textwidth]{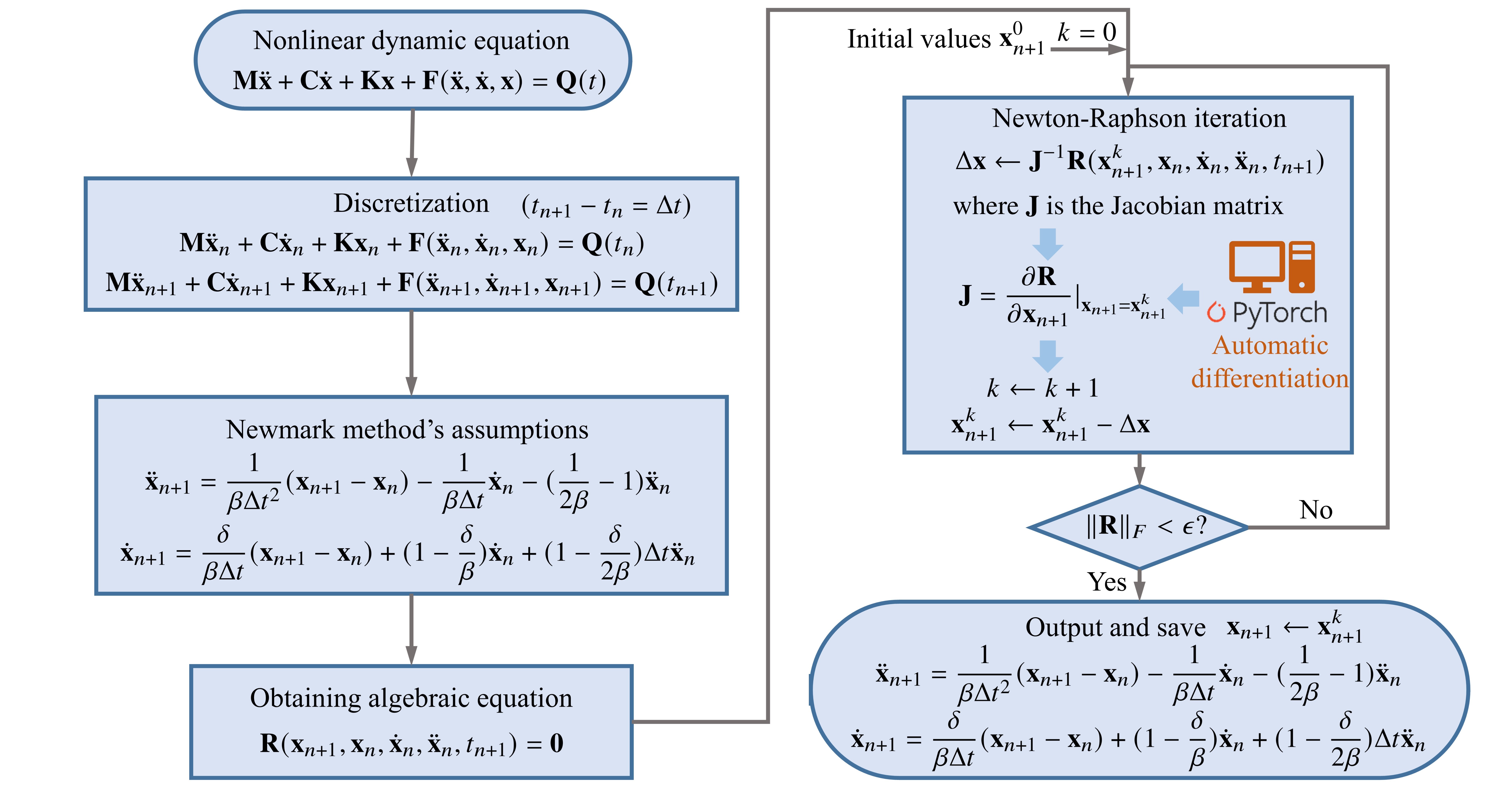}
	\caption{The computational flowchart of the modified Newmark/Newton-Raphson method with AD method for solving nonlinear dynamic systems.}
	\label{fig:new_ad}
\end{figure}

Although the traditional NNR method is algorithmically intuitive, the nonlinear form of the system we are interested in can be complicated in real-world engineering scenarios. Thus, the Jacobian matrix becomes a problem requiring additional attention. Manual differentiation in the form of symbolic calculations are usually very tedious and error-prone, especially when the system has nonlinear damping or mass. Moreover, this paradigm also makes it challenging to generalize programming. Starting from these points, we incorporate AD to obtain the Jacobian matrix required for Newton iterations, thereby establishing the modified Newmark/Newton-Raphson method with AD (NNR-AD), as illustrated in Fig. \ref{fig:new_ad}. 

Specifically, the application of automatic differentiation on Eq. (\ref{equ:equ7}) using the PyTorch \cite{ansel_pytorch_2024} is straightforward, as we show in Listing \ref{lst:code1}. Note that only some key sections are presented here for brevity. Entire project codes and more details are provided in \url{https://github.com/shuizidesu/nnr-ad}. Firstly, by carefully organizing the program's structure, we automatically acquire the algebraic equation for the NNR method (shown in Eq.\eqref{equ:equ7}) without any manual differentiation. As shown in lines 5-35 of Listing \ref{lst:code1}, for an arbitrary nonlinear system, we only need to specify the specific values for $\mathbf{M}$, $\mathbf{C}$, and $\mathbf{K}$, the form of the given external force $\mathbf{Q}(t)$ and nonlinear term, then, we can directly obtain the algebraic equations with only $\mathbf{x}_{n+1}$ unknown in a very intuitive route, i.e., by subtracting the left and right sides of Eq.\eqref{equ:equ3} (illustrated in lines 15-19 and 32-35 of Listing \ref{lst:code1}). Furthermore, based on this, we can easily obtain the Jacobian matrix using the automatic differentiation technique provided by PyTorch, specifically through the $jacrev$ function, as shown in lines 39-40 of Listing \ref{lst:code1}, which is an advanced function introduced in PyTorch version 1.13.0 that is essentially an efficient vectorization of the reverse mode automatic differentiation (because $m=n$ in the considered problem). This treatment of the Jacobian matrix is equally applicable to nonlinear systems with arbitrary forms and does not depend on tedious symbolic computation. In summary, our proposed NNR-AD method provides a completely general and very easy-to-use toolbox for obtaining the dynamical response of complex nonlinear systems through a well-structured program design and the introduction of automatic differentiation.

\begin{lstlisting}[caption={Some key parts of the code implementation of the proposed modified Newmark/Newton-Raphson method with AD.}, label={lst:code1}]
      '''Import the required libraries.'''
      import torch
      from torch.func import jacrev
      ...

      '''Define the Newmark-beta method'''
      class NewmarkBetaMethod(torch.nn.Module):
            def __init__(self, gamma=0.5, delta=0.25, delta_t=0.01):
                  ...
            def initialize_system(self, system):
                  ...
            def calculate_ddx1(self, x0, dx0, ddx0, x1):
                  ...
            def calculate_dx1(self, dx0, ddx0, ddx1):
                  ...
            def calculate_residual(self, x1, x0 ,dx0, ddx0, current_time):
                  ddx1 = self.calculate_ddx1(x0, dx0, ddx0, x1)
                  dx1 = self.calculate_dx1(dx0, ddx0, ddx1)
                  residual_equation = self.system.calculate_residual(x1, dx1, ddx1, current_time)
                  return residual_equation
                  
      '''Define the system equation'''
      class MySystem(torch.nn.Module):
            def __init__(self):
                  ...
            def initialize_parameters(self, parameters):
                  ...
            def force(self, current_time):
                  ...
            def nonlinearity(self, x, dx, ddx, current_time):
                  ...
            def prepare_initial_ddx0(self, x0, dx0, parameters, current_time):
                  ...
            def calculate_residual(self, x, dx, ddx, current_time):
                  Qt = self.force(current_time)
                  residual_vector = self.M @ ddx + self.C @ dx + self.K @ x + self.nonlinearity(x, dx, ddx, current_time) - Qt
                  return residual_vector

      ...

      '''Solving and performing Newton iterations'''
      ...
                  # Computing the Jacobi matrix using automatic differentiation (reverse mode) through the jacrev function
                  jacobian = jacrev(numerical_solver.calculate_residual, argnums=0)(x1, x0, dx0, ddx0, current_time).squeeze()
                  # Computing the residual vector
                  residual_equation = numerical_solver.calculate_residual(x1, x0, dx0, ddx0, current_time)
                  # Solving the linear equations and computing the increment
                  delta_x1 = torch.linalg.solve(jacobian, residual_equation)
                  # Updating the solution
                  x1 = x1-delta_x1
\end{lstlisting}

\section{Numerical examples}
\label{sec:num}
To demonstrate the generality and accuracy of the NNR-AD method in solving complex nonlinear systems, four representative numerical examples are presented here. The first example involves typical nonlinear systems, including the van der Pol system, Duffing system, and nonlinear pendulum system. The second example considers a 40 degrees of freedom system with complex nonlinear stiffness. The third example involves a more challenging system that simultaneously possesses nonlinear stiffness and nonlinear damping with highly complex nonlinear characteristics. The fourth example involves a high-dimensional nonlinear stiffness system with 284 degrees of freedom.

\subsection{Typical nonlinear systems}
\begin{figure}[pos=htbp]
	\centering
	\includegraphics[width=1.0\textwidth]{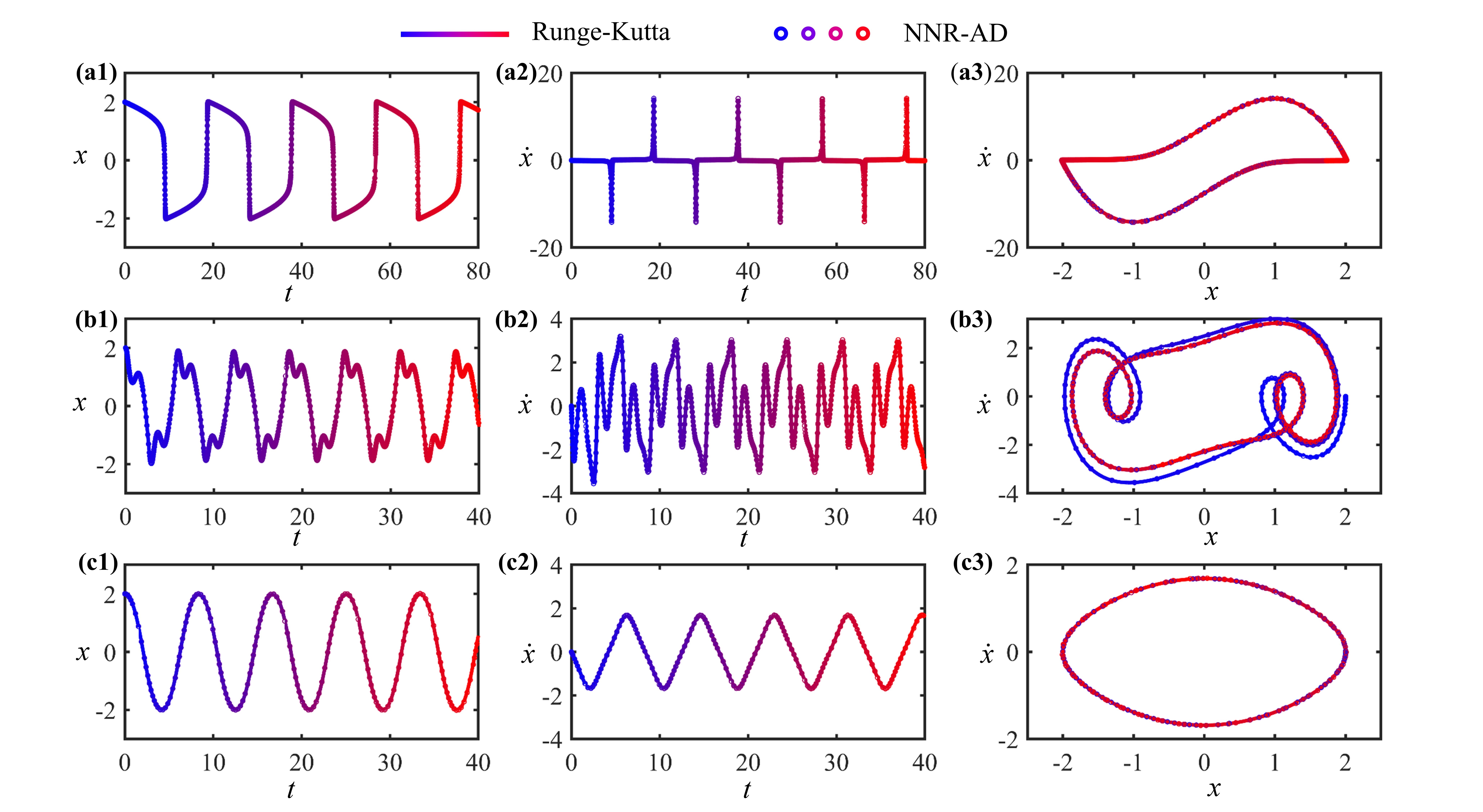}
	\caption{The displacement time histories, velocity time histories, and phase trajectories of the nonlinear systems obtained through both NNR-AD and Runge-Kutta methods. (\textbf{a}) van der Pol system. (\textbf{b}) Duffing system. (\textbf{c}) Nonlinear pendulum system. (\textbf{a1-c1}) Displacement time histories of the systems. (\textbf{a2-c2}) Velocity time histories of the systems. (\textbf{a3-c3}) Phase trajectories of the systems.}
	\label{fig:nonlinear1}
\end{figure}

The nonlinear dynamic model of the van der Pol system is given as follows:
\begin{equation}
      \ddot{x} +\varepsilon (x^2-1)\dot{x} +x=0,
\end{equation}
in the equation, $\varepsilon$ represents the parameter associated with nonlinear damping, and all results are obtained with $\varepsilon=1$. As is well known, when $\varepsilon=1$, the van der Pol oscillator exhibits self-excited oscillations and possesses a globally attracting limit cycle. The nonlinear dynamic responses of the system are numerically investigated using both NNR-AD and Runge-Kutta methods with initial conditions $x_0=2$ and $\dot{x}_0=0 $ at $t=0$. The comparative results, including displacement time history, velocity time history, and phase trajectory, are presented in Fig. \ref{fig:nonlinear1}(a). As clearly demonstrated in Fig. \ref{fig:nonlinear1}(a1), the system exhibits pronounced relaxation oscillations, with complete agreement between the computational results obtained from both methods. This remarkable consistency conclusively verifies the accuracy of the NNR-AD method when applied to the van der Pol system.

The nonlinear dynamical model of the Duffing system is governed by the following equation:
\begin{equation}
      \ddot{x} +\delta \dot{x} +\alpha x+\beta x^3=\gamma \cos(\omega t),
\end{equation}
where $\delta$, $\alpha$, $\beta$, $\gamma$ and $\omega$ are the system's parameters, and here we choose $\delta = 1$, $\alpha = 1$,  $\beta = 3$, $\gamma = 10$,and $\omega = 1$. When $t = 0$, the initial values of the given system are $x_0=2$ and $\dot{x}_0=0 $. The nonlinear dynamic model of the system is calculated using both NNR-AD and Runge-Kutta methods. The displacement time history, velocity time history and phase trajectory of the system are shown in Fig. \ref{fig:nonlinear1}(b). As shown in Fig. \ref{fig:nonlinear1}(b3), there are two basins of attraction in this system. Moreover, the displacement time history, velocity time history and phase trajectory obtained by the two calculation methods are completely coincident, which proves the accuracy of the NNR-AD method. That is to say, the NNR-AD method proposed in this paper is applicable to the Duffing system.

The nonlinear pendulum system is governed by the following dynamic equation:
\begin{equation}
      \ddot{x} +\sin(x)=0,
\end{equation}
similarly, given the initial value $x_0=2$ and $\dot{x}_0=0 $ at $t=0$, the solutions are calculated using NNR-AD and Runge-Kutta respectively, and the displacement time history, velocity time history and phase trajectory obtained are shown in Fig. \ref{fig:nonlinear1}(c). From Fig. \ref{fig:nonlinear1}(c3), it can be observed that the nonlinear pendulum system exhibits a closed orbit around a single center $(x, \dot{x})=(0,0)$. Moreover, it can be found that the response curves obtained by the two calculation methods have good consistency. In other words, the NNR-AD proposed in this paper is applicable to the nonlinear pendulum system.

\subsection{40 degrees of freedom system with complex nonlinear stiffness}
\begin{figure}[pos=htbp]
	\centering
	\includegraphics[width=1.0\textwidth]{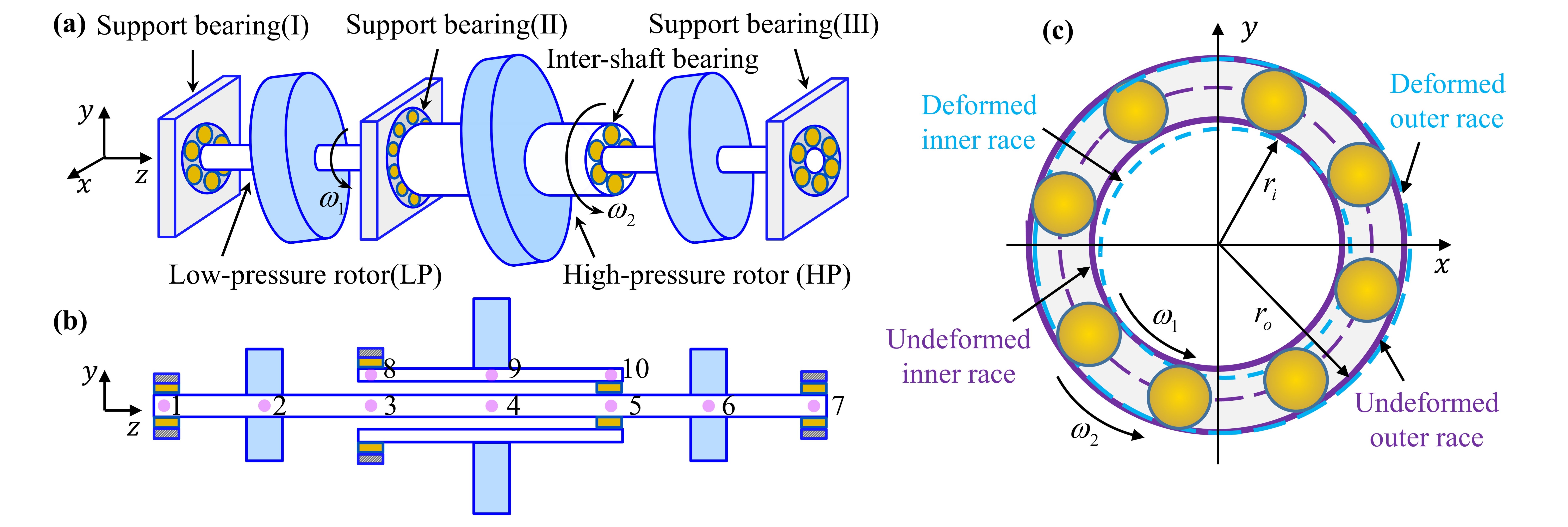}
	\caption{Schematic diagram of the nonlinear stiffness system. (\textbf{a}) Schematic diagram of the dual-rotor system. (\textbf{b}) Schematic diagram of the finite element model of the dual-rotor system. (\textbf{c}) Schematic diagram of the bearing.}
	\label{fig:nonlinear_k1}
\end{figure}

To further validate the universality and accuracy of the NNR-AD method in solving complex nonlinear dynamic systems, we analyzed a 40 degrees of freedom dual-rotor system model with multiple nonlinear stiffness supports. The schematic diagram of the model is shown in Fig. \ref{fig:nonlinear_k1}(a), which consists of a high-pressure rotor, a low-pressure rotor, support bearings at the far left and right ends of the low-pressure rotor, a support bearing at the far left end of the high-pressure rotor, and an inter-shaft bearing connecting the high- and low-pressure rotors. where the rotational speed of the low-pressure rotor is $\omega _1$, and that of the high-pressure rotor is $\omega _2$. Both the support bearings and the inter-shaft bearing employ the bearing model illustrated in Fig. \ref{fig:nonlinear_k1}(c). This model utilizes the Hertz contact theory, where the nonlinear bearing support force is determined based on the relative displacement between the inner and outer races.

To obtain the nonlinear dynamic model of the system illustrated in Fig. \ref{fig:nonlinear_k1}(a), a finite element model (see Fig. \ref{fig:nonlinear_k1}(b)) is established based on finite element theory. In this model, the low-pressure rotor is discretized into 7 nodes, while the high-pressure rotor is divided into 3 nodes. Each node possesses 4 degrees of freedom, and since the model comprises 10 nodes, the system exhibits 40 degrees of freedom in total. Nodes 2 and 6 represent the low-pressure rotor disks, while node 9 corresponds to the high-pressure rotor disk. Nodes 1, 7, and 8 are subjected to nonlinear support bearing forces, whereas nodes 5 and 10 experience nonlinear inter-shaft bearing forces.

Using the Lagrange equations of the second kind, the dynamic equations of the disks are derived as follows.
\begin{equation}
      \begin{cases}
            {\textbf{M}_d}{{\ddot{\textbf{U}}}_{d1}} - \varOmega \textbf{J}{\dot{\textbf{U}}_{d2}} = {\textbf{Q}_{d1}}\\
            {\textbf{M}_d}{{\ddot{\textbf{U}}}_{d2}} + \varOmega \textbf{J}{\dot{\textbf{U}}_{d1}} = {\textbf{Q}_{d2}}
      \end{cases},
\end{equation}
where, $\textbf{U}_{d1} = [x, - {\theta _y}]^T$, $\textbf{U}_{d2} = [y, \theta _x]^T$, and 
$\textbf{M}_d=\left[
\begin{array}
{cc}
m&0\\
0&J_d
\end{array}
\right]$ is the mass matrix of the disk, and $\textbf{J}=\left[
\begin{array}
{cc}
0&0\\
0&J_p
\end{array}
\right]$ is the inertia matrix of the disk, and $\textbf{G}_d=\varOmega \textbf{J}$ is the rotation matrix of the disk. $\textbf{Q}_{d1}$ and $\textbf{Q}_{d2}$ represent the generalized forces acting on the disk in the $x$ and $y$ directions respectively. Considering the imbalance on the disk, the generalized forces acting on the disk refer to the unbalanced forces causing rotational motion on the disk, i.e. $\textbf{Q}_{d1}=[m\varOmega ^2e\sin (\varOmega t+\phi ),0]^T$, and $\textbf{Q}_{d2}=[m\varOmega ^2e\cos (\varOmega t+\phi ),0]^T$. Where, $x$ is displacement of the disk in the $x$-direction, $y$ is displacement of the disk in the $y$-direction, $\theta _x$ is rotation angle of the disk around the $x$-axis, $\theta _y$ is rotation angle of the disk around the $y$-axis, $m$ is mass of the disk, $J_d$ is diameter rotational inertia of the disk, $J_p$ is polar rotational inertia of the disk, $e$ is eccentricity of the disk, $\varOmega$ is rotational speed of the disk, $t$ is time, $\phi$ and is phase angle of the disk.

Using the Lagrange equations of the second kind, the dynamic equations of the shafts are derived as follows.
\begin{equation}
      \begin{cases}
            {\textbf{M}_s}{\ddot{\textbf{U}}_{s1}} - \varOmega \textbf{J}_s{\dot{\textbf{U}}_{s2}} + {\textbf{K}_s}{\textbf{U}_{s1}}= {\textbf{Q}_{s1}}\\
            {\textbf{M}_s}{\ddot{\textbf{U}}_{s2}} + \varOmega \textbf{J}_s{\dot{\textbf{U}}_{s1}} + {\textbf{K}_s}{\textbf{U}_{s2}}= {\textbf{Q}_{s2}}
      \end{cases},
\end{equation}
where, $\textbf{U}_{s1} = [x_1, - {\theta _{y1}},x_2, - {\theta _{y2}}]^T$, and $\textbf{U}_{s2} = [y_1, {\theta _{x1}},y_2, {\theta _{x2}}]^T$, $x_1, y_1,\theta _{x1}$ and $\theta _{y1}$ represent respectively the displacement of the first node of the shaft along the $x$-axis, along the $y$-axis, the rotation angle around the $x$-axis, and the rotation angle around the $y$-axis. $x_2, y_2,\theta _{x2}$ and $\theta _{y2}$ represent respectively the displacement of the second node of the shaft along the $x$-axis, along the $y$-axis, the rotation angle around the $x$-axis, and the rotation angle around the $y$-axis, $\varOmega$ is rotational speed of the shaft, $\textbf{M}_s$ is the mass matrix of the shaft, $\textbf{J}_s$ is the inertia matrix of the shaft,and $\textbf{K}_s$ is the stiffness matrix of the shaft. The detailed information on the above matrix is shown in the Appendix \ref{app:A}. $\textbf{Q}_{s1}$ and $\textbf{Q}_{s2}$ represent the nonlinear bearing support forces acting on the shaft in the $x$ and $y$ directions respectively. 

The bearing (see Fig. \ref{fig:nonlinear_k1}(c)) contains many nonlinearities, such as the fractional exponential of Hertzian contact, the varying stiffness excitation and the radial clearance. The rotational speed of the bearing outer ring is $\omega _2$, while the inner ring rotates at $\omega _1$, resulting in a cage speed of $\omega _c=\frac{r_i\omega _2+r_o\omega _1}{r_i+r_o} $, where $r_i$ represents the inner race radius and $r_o$ denotes the outer race radius. For the inter-shaft bearing, the inner and outer ring speeds correspond to the low-pressure and high-pressure rotor speeds, respectively. Regarding the low-pressure and high-pressure support bearings, the outer ring speed remains stationary, with the inner ring rotating at the low-pressure rotor speed and high-pressure rotor speed, respectively. The angular position of the $k$-th rolling element in the bearing at any given time $t$ can be expressed as:
\begin{equation}
      \theta _k=\frac{2\pi }{N_b} (k-1)+\omega _ct(k=1,2,...,N_b),
\end{equation}
where, $N_b$ is the number of rolling elements.

The relative contact deformation $\delta _k$ of the $k$-th rolling element with respect to the bearing races can be expressed as:
\begin{equation}
      \delta _k=(x_i-x_o)\cos \theta _k + (y_i-y_o)\sin \theta _k - \delta _0,
\end{equation}
where, $x_i,x_o,y_i,y_o$ represent the relative displacements between the bearing inner and outer races in the $x$ and $y$ directions, $\delta _0$ is the radial clearance of the bearing.

The bearing support force can be obtained using the Hertz contact model as follows:
\begin{equation}
      \left[\begin{array}{c}F_x\\F_y\\\end{array}\right]=K_b\sum_{k = 1}^{N_b}{\delta _k}^n H(\delta _k) \left[\begin{array}{c}\cos(\theta _k)\\\sin(\theta_k)\\\end{array}\right],
      \label{equ:f_bear}
\end{equation}
where $H(i) = \begin{cases}1(i>0)\\0(i\leqslant 0)\end{cases}$, $K_b$ represents the Hertz contact stiffness of the bearing, and $n=\frac{10}{9} $.

According to the finite element method, by assembling the disk elements, shaft elements, and nonlinear bearing supports, the mass matrix, inertia matrix, stiffness matrix, and external force matrix of the system are obtained. The nonlinear dynamic equations of the system are as follows.
\begin{equation}
      \textbf{M}\ddot{\textbf{X}}+\textbf{C}\dot{\textbf{X}}+\textbf{KX}+\textbf{F}_{nonlinear}(\textbf{X})=\textbf{F}(t),
      \label{equ:k}
\end{equation}
where, $\textbf{M}$ is the mass matrix of the system, $\textbf{C}$ is composed of the system's damping matrix and inertia matrix, $\textbf{K}$ is the system's stiffness matrix, $\textbf{F}_{nonlinear}(\textbf{X})$ represents the nonlinear bearing support force acting on the system, as can be seen from Eq. (\ref{equ:f_bear}), $\textbf{F}_{nonlinear}(\textbf{X})$ is solely a nonlinear expression of $\textbf{X}$, thus we define this system as a nonlinear stiffness model, $\textbf{F}(t)$ denotes the unbalance force of the rotor disk and gravitational force of the system, $\textbf{X}=[x_1,y_1,\theta _{x1},\theta _{y1},...,x_{10},y_{10},\theta _{x10},\theta _{y10}]$.

\begin{figure}[pos=htbp]
	\centering
	\includegraphics[width=1.0\textwidth]{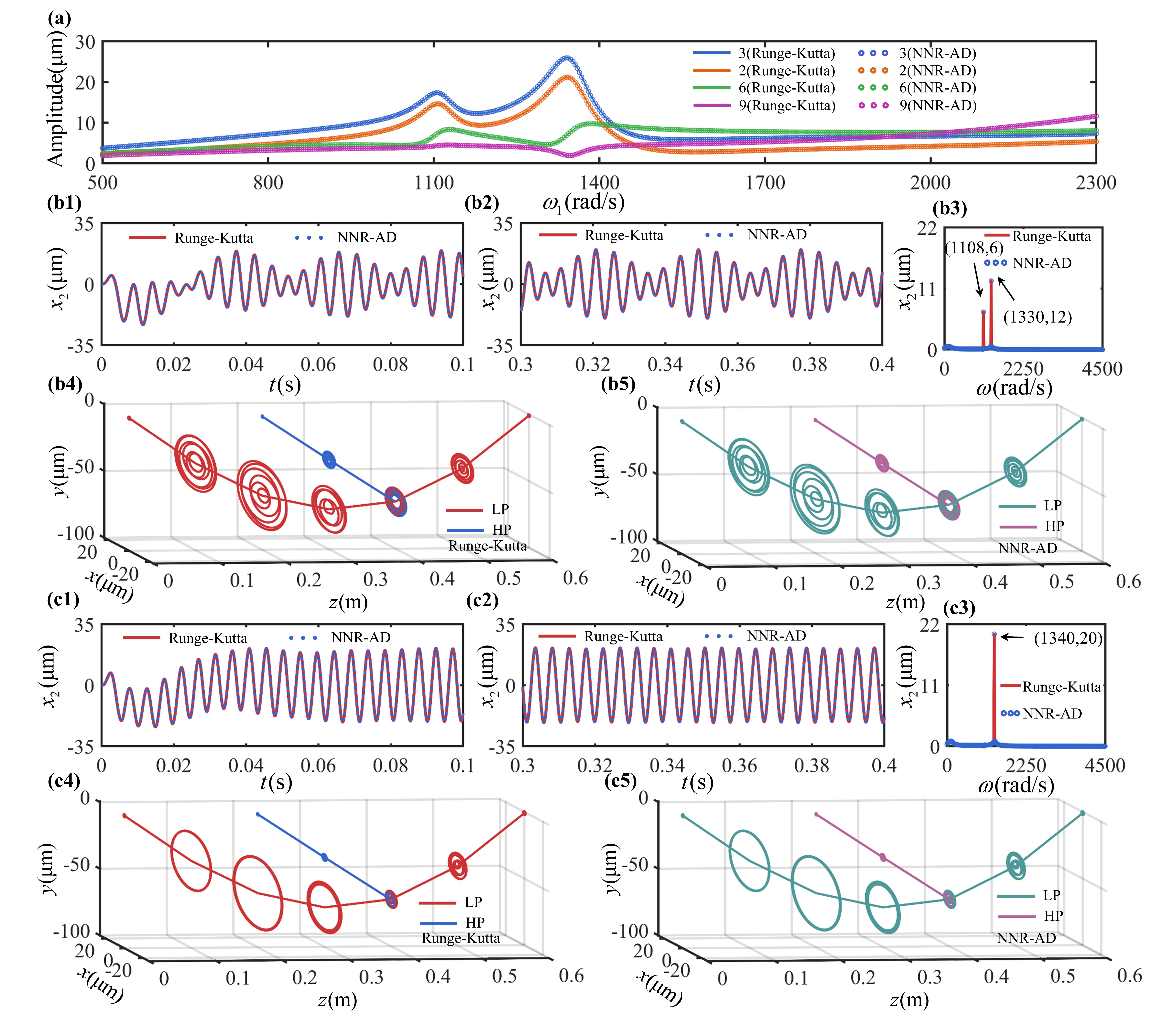}
	\caption{The dynamic responses of the system obtained through NNR-AD and Runge-Kutta methods. (\textbf{a}) Amplitude-frequency response of the dual-rotor system (stable phase). (\textbf{b}) The time history, frequency spectrum, and rotor's orbit of the dual-rotor system at $\omega _1=1108$rad/s and $\omega _2=1330$rad/s. (\textbf{c}) The time history, frequency spectrum, and rotor's orbit of the dual-rotor system at $\omega _1=1340$rad/s and $\omega _2=1608$rad/s. (\textbf{b1,c1}) The initial time history of $x$-direction displacement at Node 2. (\textbf{b2,c2}) The time history of $x$-direction displacement at Node 2 during the stable phase. (\textbf{b3,c3}) The frequency spectrum of the $x$-direction displacement at Node 2 (stable phase). (\textbf{b4,c4}) The rotor's orbit of the system obtained by Runge-Kutta (stable phase). (\textbf{b5,c5}) The rotor's orbit of the system obtained by NNR-AD (stable phase).}
	\label{fig:nonlinear_k}
\end{figure}

The nonlinear Eq. (\ref{equ:k}) is solved using both NNR-AD and Runge-Kutta methods, respectively, and the obtained dynamic responses are shown in Fig. \ref{fig:nonlinear_k}. Considering that Eq. (\ref{equ:k}) contains displacements in both $x$ and $y$ directions for the high- and low-pressure rotors, and the responses obtained through numerical methods are all discrete, the following equation is adopted to define the system amplitude at a given rotational speed, which comprehensively characterizes the vibration amplitude of each node in the dual-rotor system during the stable phase:
\begin{equation}
      A=\sqrt{\frac{\sum_{i = 1}^{N}(((x(i)-\overline{x} )^2+(y(i)-\overline{y} )^2))}{N} } ,
      \label{equ:A}
\end{equation}
where, $A$ represents the vibration amplitude, $N$ represents the number of discrete points, $x$ denotes the discrete displacement response in the $x$-direction, $y$ signifies the discrete displacement response in the $y$-direction, $\overline{x}$ stands for the mean displacement response in the $x$-direction, and $\overline{y} $ indicates the mean displacement response in the $y$-direction. The amplitude-frequency response of the dual-rotor system calculated by both Runge-Kutta and NNR-AD methods are shown in Fig. \ref{fig:nonlinear_k}(a), with $\omega_1$ ranging from 500 rad/s to 2300 rad/s and $\omega_2$ maintained at 1.2 times the low-pressure speed (600 rad/s - 2760 rad/s). The results demonstrate complete agreement between the two computational methods regarding the amplitude-frequency response at nodes 3, 2, 6, and 9, confirming the accuracy of the NNR-AD method. Furthermore, resonance phenomena are observed at the low-pressure rotor nodes 2 and 3 when the $\omega_1$ reach 1108 rad/s and 1340 rad/s, respectively. Consequently, additional analyses are conducted for two specific operational conditions: (1) $\omega_1=1108$rad/s and $\omega_2=1330$rad/s, (2) $\omega_1=1340$rad/s and $\omega_2=1608$rad/s. The corresponding dynamic responses obtained from both computational methods are presented in Fig. \ref{fig:nonlinear_k}(b) and (c). 

Considering the prolonged computational duration required for all system nodes, the displacement time histories from $t=0$ to the stable phase appear excessively dense in the plot, which hinders direct visual comparison of the results' accuracy between the two computational methods. Therefore, we separately extract two segments of the $x_2$ displacement time history: the initial phase and the stable phase, as shown in Fig. \ref{fig:nonlinear_k}(b1) and (b2) respectively. The results demonstrate that both computational methods yield identical outcomes, regardless of whether the system is in the initial response phase or has reached stable operation. As evidenced by Fig. \ref{fig:nonlinear_k}(b3), the frequency spectrum obtained through both computational methods demonstrate complete consistency. The system frequencies in this case contain both the low-pressure rotational speed (1108 rad/s) and high-pressure rotational speed (1330 rad/s), with the high-pressure rotational speed component being dominant. Furthermore, the rotor's orbits obtained through Runge-Kutta (Fig. \ref{fig:nonlinear_k}(b4)) and NNR-AD (Fig. \ref{fig:nonlinear_k}(b5)) are plotted separately. The results show complete consistency between both rotor's orbits, with Node 3 exhibiting the largest orbital range, thereby further validating the accuracy of the NNR-AD. For Fig. \ref{fig:nonlinear_k}(c), similarly, the displacement time histories (both initial and stable phases) and frequency spectra of $x_2$ obtained by both computational methods show perfect agreement. Unlike the case at rotational speed $\omega_1=1108$rad/s and $\omega_2=1330$rad/s, the system's response frequencies in this scenario contain only the low-pressure speed (1340 rad/s), without high-pressure speed (1608 rad/s). Moreover, the rotor's orbits computed by both methods remain completely consistent, with Node 3 still exhibiting the largest orbital range among all system nodes. Through the analysis of Fig. \ref{fig:nonlinear_k}, it can be conclusively demonstrated that the NNR-AD is fully applicable to high-degree-of-freedom dynamic systems incorporating complex nonlinear stiffness characteristics.

\subsection{System with highly complex nonlinear stiffness and damping}

\begin{figure}[pos=htbp]
	\centering
	\includegraphics[width=0.96\textwidth]{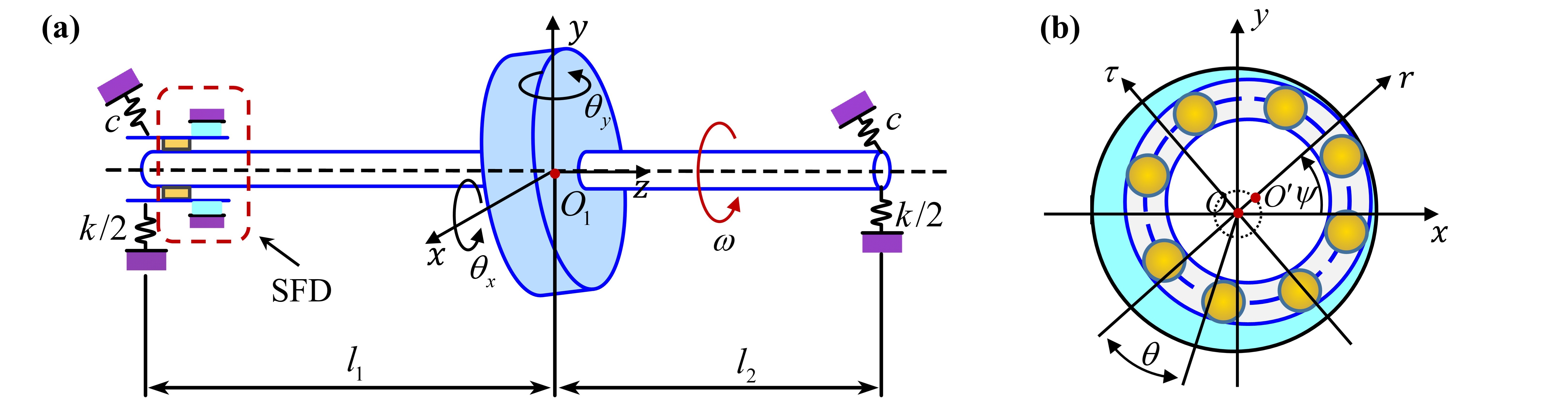}
	\caption{Schematic diagram of the nonlinear stiffness and damping system. (\textbf{a}) Schematic diagram of the rotor system with SFD. (\textbf{b}) Schematic diagram of the SFD.}
	\label{fig:nonlinear_c1}
\end{figure}

To further validate the universality and accuracy of the NNR-AD method in solving complex nonlinear dynamic systems, we investigate a rotor system with complex nonlinear stiffness and damping characteristics. The rotor model is supported by a squeeze film damper (SFD) on the left side and a conventional linear bearing on the right side, where the SFD represents a typical system exhibiting complex nonlinear damping and stiffness characteristics. The schematic diagram of the rotor model with SFD is shown in Fig. \ref{fig:nonlinear_c1}(a), and the schematic diagram of the SFD is shown in Fig. \ref{fig:nonlinear_c1}(b).

According to the second kind of Lagrange's equations, the nonlinear dynamic equations for the rotor system with SFD shown in Fig. \ref{fig:nonlinear_c1}(a) can be derived as follows:
\begin{equation}
      \begin{cases}
            m\ddot{x} + 2c\dot{x} + c(l_1-l_2)\dot{\theta }_y + kx +\frac{k(l_1-l_2)}{2}\theta _y + F_x = \delta \omega ^2 \cos (\omega t)\\
            m\ddot{y} + 2c\dot{y} + c(l_2-l_1)\dot{\theta }_x + ky +\frac{k(l_2-l_1)}{2}\theta _x + F_y = \delta \omega ^2 \sin (\omega t)\\
            J_d\ddot{\theta }_x + c(l_2-l_1)\dot{y} + c({l_1}^2+{l_2}^2)\dot{\theta}_x + J_p\omega\dot{\theta}_y +\frac{k(l_2-l_1)}{2}y + \frac{k({l_1}^2+{l_2}^2)}{2}\theta _x  - F_yl_1 = 0\\
            J_d\ddot{\theta }_y + c(l_1-l_2)\dot{x} + c({l_1}^2+{l_2}^2)\dot{\theta}_y - J_p\omega\dot{\theta}_x +\frac{k(l_1-l_2)}{2}x + \frac{k({l_1}^2+{l_2}^2)}{2}\theta _y  + F_xl_1 = 0
            \label{equ:c}
      \end{cases},
\end{equation}
where, $x, y$ represent the translational displacements of the disk center $O _1$ along the $x$- and $y$-axes, $\theta _x, \theta _y$ denote the rotational angles about the $x$- and $y$-axes, $m$ is the mass of the disk, $J_d$ and $J_p$ are the diametral and polar moments of inertia of the disk, $c$ is the damping coefficient of the linear support, $k$ is the stiffness coefficient of the linear support, $\omega $ is the rotational speed of the rotor, $\delta $ is the unbalance magnitude of the disk, $l_1$ and $l_2$ are the lengths of the shaft, $t$ is the time, $F_x$ and $F_y$ are the horizontal and vertical components of the nonlinear oil film force provided by the SFD, respectively.

The cross-section of SFD is illustrated in the rotating coordinate system in Fig. \ref{fig:nonlinear_c1}(b), where the journal center $O '$ orbits around the geometric center $O $ of the oil film outer ring. Defining $\theta $ from the position of maximum oil film thickness, and for SFD with length-to-diameter ratio $L/D < 0.25$ without end seal, the radial and tangential oil film forces under the short-bearing assumption can be expressed as:
\begin{equation}
      \begin{cases}
            F_r = \frac{\mu R L^3}{C^2}({I_3}^{11}\dot{\psi }r + {I_3}^{02}\dot{r}) \\
            F_t = \frac{\mu R L^3}{C^2}({I_3}^{20}\dot{\psi }r + {I_3}^{11}\dot{r}) \\
      \end{cases},
\end{equation}
where, $\mu$ represents the lubricant viscosity, $R$ is the journal radius, $C$ denotes the oil film clearance, $r=e/C$ represents the dimensionless radial displacement, $e$ is defined as the radial displacement of the journal center, expressed as:
\begin{equation}
      e = \sqrt{(x+\theta _yl_1)^2 + (y-\theta _xl_1)^2},
\end{equation}
$\psi$ is defined as the precession angle of the journal, given by:
\begin{equation}
      \psi =\arctan\frac{y-\theta _xl_1}{x+\theta _yl_1},
\end{equation}
$I$ represents the Sommerfeld integral defined as follows:
\begin{equation}
{I_3}^{lm} = \int_{\theta _1}^{\theta _2} {\frac{\sin^l\theta \cos^m\theta   }{(1+r\cos\theta)^3} } \,d\theta  (l,m=0\sim 2), 
\end{equation}
where,$\theta _1$ and $\theta _2$ denote the starting and ending points of the positive pressure zone in the oil film, respectively, $\theta _1 = \arctan -(\frac{\dot{r} }{r\dot{\psi}}) $, $\theta _2 = \theta _1+\pi $. 

Considering that the numerical solution of Eq. (\ref{equ:c}) involves Sommerfeld integration, which is a nonlinear function of $(x, y, \theta _x, \theta _y,\dot{x} ,\dot{y},\dot{\theta }_x,\dot{\theta }_y)$, we employ Gaussian-Legendre quadrature to transform the analytical form of Sommerfeld integration into a numerical integration format with 15 Gaussian nodes, achieving 29th-order algebraic precision, as expressed below:
\begin{equation}
      {I_3}^{lm}\approx \frac{\theta _2-\theta _1}{2}  \sum_{i = 1}^{15}  w_i{\frac{\sin^l(\frac{\theta _2-\theta _1}{2}t_i+ \frac{\theta _2+\theta _1}{2}) \cos^m(\frac{\theta _2-\theta _1}{2}t_i+ \frac{\theta _2+\theta _1}{2})}{(1+r\cos(\frac{\theta _2-\theta _1}{2}t_i+ \frac{\theta _2+\theta _1}{2}))^3} }(l,m=0\sim 2),
\end{equation}
where, $t_i$ are the roots of the 15th-order Legendre polynomial (Gauss nodes), $w_i$ are the corresponding quadrature weights.

The transformation of oil film forces from the rotating coordinate system to the Cartesian coordinate system is given by:
\begin{equation}
      \begin{cases}
            F_x = F_r\frac{x+\theta _yl_1}{e} - F_t\frac{y-\theta _xl_1}{e} \\
            F_y = F_r\frac{y-\theta _xl_1}{e} + F_t\frac{x+\theta _yl_1}{e} 
      \end{cases},
\end{equation}
it can be observed that both $F_x$ and $F_y$ exhibit an extremely complex nonlinear mapping relationship with $(x, y, \theta _x, \theta _y,\dot{x} ,\dot{y},\dot{\theta }_x,\dot{\theta }_y)$. Consequently, Eq. (\ref{equ:c}) incorporates highly intricate nonlinear stiffness and damping forces, making it a dynamic equation characterized by exceptionally complex nonlinear stiffness and damping characteristics. Obtaining the Jacobian matrix for this equation proves exceptionally challenging through manual differentiation—not only is the process labor-intensive and time-consuming, but it also cannot guarantee correctness. Furthermore, symbolic differentiation often leads to expression swell, significantly complicating the computational implementation.

The physical parameters of the system are as follows: $m=37.62$ kg, $k=5.4\times 10^6$N/m, $J_d=0.8$ kg$\cdot $m$^2$, $J_p=1.6$ kg$\cdot $m$^2$, $l_1=0.894$m, $l_2=1.038$m, $c=265$N$\cdot $s/m, $\delta =6.508\times 10^{-4}$kg$\cdot $m, $C=2.5\times 10^{-4}$m, $R=3.915\times 10^{-2}$m, $L=0.015$m, $\mu =6.76\times 10^{-3}$N$\cdot $s/m.

\begin{figure}[pos=htbp]
	\centering
	\includegraphics[width=0.99\textwidth]{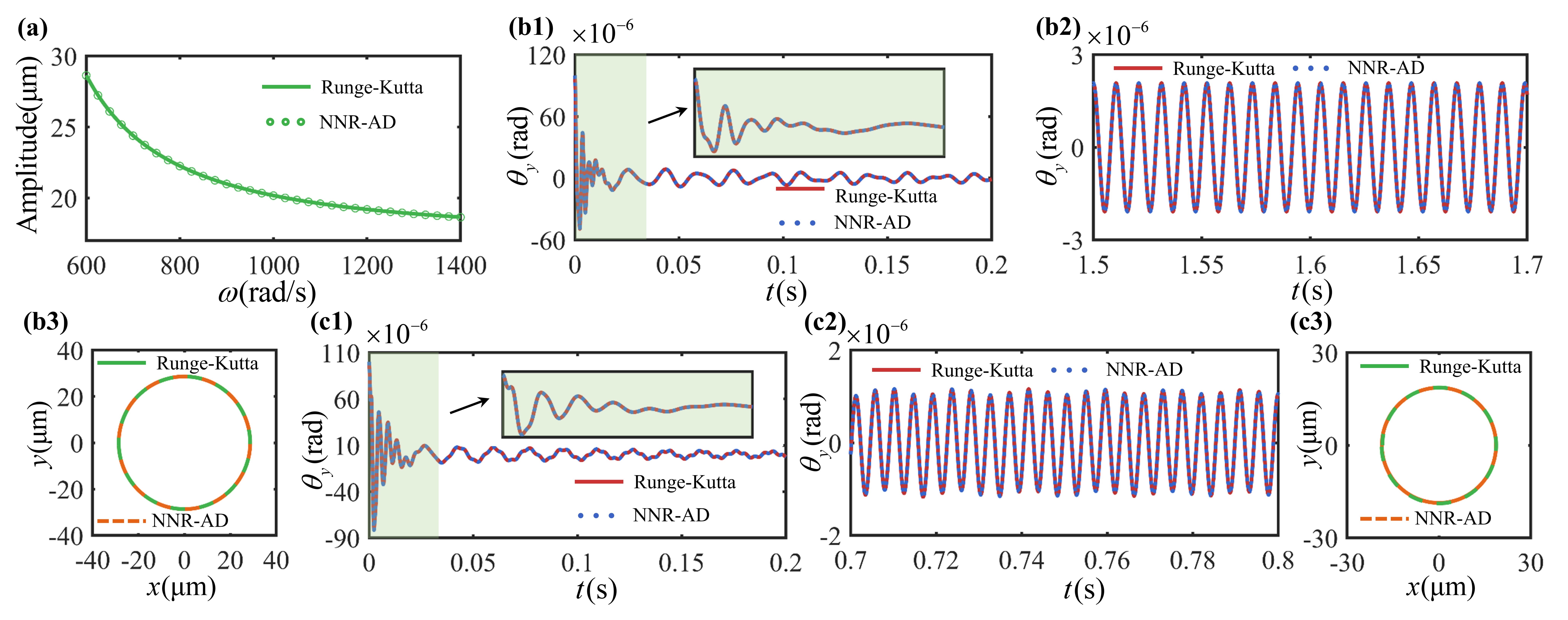}
	\caption{The dynamic responses of the system obtained through NNR-AD and Runge-Kutta methods. (\textbf{a}) Amplitude-frequency response of the rotor system with SFD (stable phase). (\textbf{b}) The time history, and rotor's orbit of the rotor system with SFD at $\omega =600$rad/s . (\textbf{c}) The time history, and rotor's orbit of the rotor system with SFD at $\omega =1400$rad/s. (\textbf{b1,c1}) The initial time history of $\theta _y$. (\textbf{b2,c2}) The time history of $\theta_y$ during the stable phase. (\textbf{b3,c3}) The rotor's orbit of the rotor system with SFD (stable phase).}
	\label{fig:nonlinear_c}
\end{figure}

The nonlinear Eq. (\ref{equ:c}) is solved using both NNR-AD and Runge-Kutta methods, respectively, and the obtained dynamic responses are shown in Fig. \ref{fig:nonlinear_c}. Using the definition of system amplitude given by Eq. (\ref{equ:A}), the amplitude-frequency response of the rotor system with SFD is plotted over the rotational speed range of 600 rad/s to 1400 rad/s, as shown in Fig. \ref{fig:nonlinear_c}(a). The results demonstrate that both computational methods yield identical amplitude-frequency responses. Furthermore, it can be observed that the system amplitude decreases with increasing rotational speed. 

At $\omega $ = 600 rad/s, the initial time history of $\theta_y$ is shown in Fig. \ref{fig:nonlinear_c}(b1). It can be observed that the results obtained by both computational methods exhibit excellent agreement, with complete consistency even during the initial calculation phase. When the computation time is extended, the time history presented in Fig. \ref{fig:nonlinear_c}(b2) demonstrates that the results from both methods remain consistent. Moreover, the time history curves exhibit periodic behavior, indicating that the system has reached a steady state. The corresponding rotor's orbit, plotted in Fig. \ref{fig:nonlinear_c}(b3), confirms that both methods yield identical circular trajectories.
A further investigation is conducted at $\omega $ = 1400 rad/s, where the time history of $\theta_y$ during the initial phase (See Fig. \ref{fig:nonlinear_c}(c1)) and the stable phase (See Fig. \ref{fig:nonlinear_c}(c2)), as well as the corresponding rotor's orbit in the stable phase (See Fig. \ref{fig:nonlinear_c}(c3)), are plotted for both computational methods. As evident from the results, the dynamic responses obtained from the two methods exhibit perfect agreement. Through the analysis of Fig. \ref{fig:nonlinear_c}, it can be conclusively demonstrated that the NNR-AD method is fully applicable to dynamic systems with complex nonlinear stiffness and damping characteristics.

\subsection{High-dimensional system with nonlinear stiffness}
\begin{figure}[pos=htbp]
	\centering
	\includegraphics[width=0.9\textwidth]{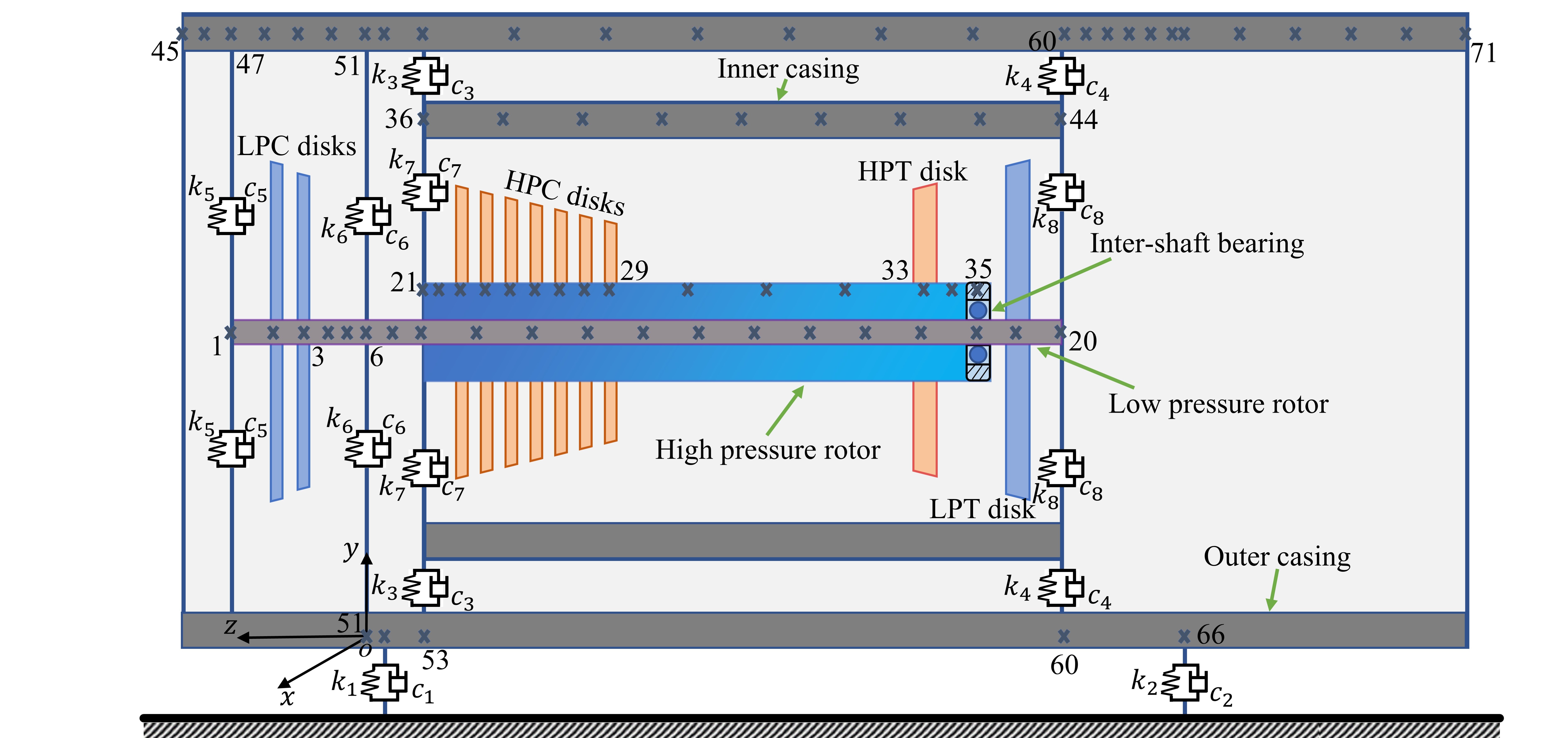}
	\caption{The schematic diagram of the 284 degrees of freedom aero-engine system with a nonlinear inter-shaft bearing.}
	\label{fig:nonlinear_284}
\end{figure}

To further validate the applicability of the NNR-AD method for high-dimensional nonlinear dynamic systems, we investigated a 284 degrees of freedom aero-engine system with a nonlinear inter-shaft bearing from Ref. \cite{chenNonlinearDynamicsAnalysis2023} (see Fig. \ref{fig:nonlinear_284}). The system comprises a low-pressure rotor, a high-pressure rotor, a nonlinear inter-shaft bearing connecting the two rotors, an inner casing, an outer casing, and additional linear support components. The nonlinear dynamic equations of this system can likewise be expressed in the form of Eq. (\ref{equ:k}) , with the distinction that $\textbf{X}$ is now defined as follows:
\begin{equation}
\textbf{X}=[x_1,y_1,\theta _{x1},\theta _{y1},x_2,y_2,\theta _{x2},\theta _{y2},...,x_{284},y_{284},\theta _{x284},\theta _{y284}].
\end{equation}

\begin{figure}[pos=htbp]
	\centering
	\includegraphics[width=0.95\textwidth]{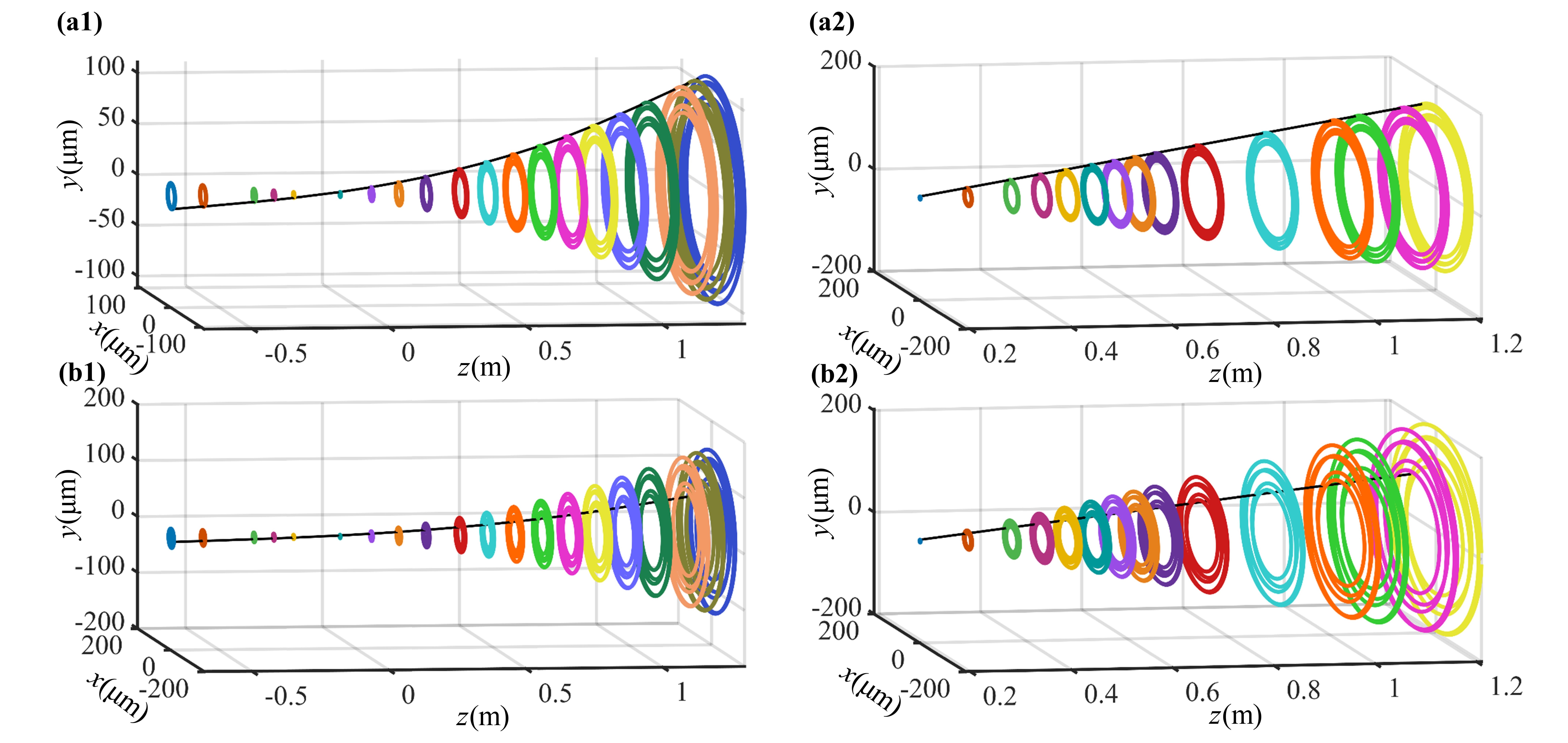}
	\caption{The rotor's orbits of the system computered by the NNR-AD method. (\textbf{a1,a2}) $\omega _1=150$rad/s and $\omega _2=180$rad/s. (\textbf{b1,b2}) $\omega _1=180$rad/s and $\omega _2=216$rad/s. (\textbf{a1,b1}) The low pressure rotor. (\textbf{a2,b2}) The high pressure rotor.}
	\label{fig:nonlinear_284_1}
\end{figure}

To validate the accuracy of the NNR-AD method in computing the dynamic responses of this system, we follow Ref. \cite{chenNonlinearDynamicsAnalysis2023} and calculate the system's dynamic responses under two operational conditions: ($\omega _1=150$rad/s , $\omega _2=180$rad/s) and ($\omega _1=180$rad/s , $\omega _2=216$rad/s). The corresponding rotor's orbits for both the low-pressure and high-pressure rotors are plotted in Fig. \ref{fig:nonlinear_284_1}. It should be noted that as this system represents a typical high-dimensional stiff system, the Runge-Kutta method employed in previous case studies failed to yield solutions, and thus only the results from the NNR-AD method are included in this section. However, we compared the rotor's orbits computed by the NNR-AD method with those obtained from the modified HB-AFT method in Ref. \cite{chenNonlinearDynamicsAnalysis2023}, and the results from both methods show excellent agreement, thereby demonstrating the accuracy of the NNR-AD method.

\section{Disscussions}
\label{sec:dic}
It should be emphasized that our proposed NNR-AD method is not mathematically different from the classical NNR method. Its advancement lies mainly in the innovative solution to the problem of calculating the necessary Jacobi matrix in a more general and convenient paradigm by combining the AD approach. In the specific program implementation, our proposed method eliminates all manual derivation processes while accurately calculating the Jacobian matrix, perfectly addressing the respective limitations in the manual differentiation, symbolic differentiation, and numerical differentiation methods.

It must be recognized, however, that the integration of AD also introduces some inebitable computational efficiency degradation since each Newtonian iteration requires a recalculation of the Jacobian matrix rather than just substituting different values, as is the case with manual and symbolic differentiation. But this is not a serious issue, as our proposed method can be run on GPUs very easily, benefiting from the excellent parallelism and CUDA support provided by advanced deep learning frameworks, such as PyTorch and JAX. Thus, our approach is still feasible even when dealing with large-scale problems, as it can fully utilize the provided computational resources. Additionally, unlike training a neural network, where a single-precision floating-point number is used as the default data type, double-precision is recommended by default in our approach because, in many practical engineering systems, the magnitude of the dynamical response is inherently small (near or below order $10^{-6}$); hence single-precision is not sufficient. This will waste some computational resources, as most GPUs are additionally optimized for single-precision floating-point number operation, which can be solved by dimensionless technique. Here we recommend double-precision by default for ease-of-use consideration. To summarize, our proposed approach achieves a significant trade-off between generality, ease-of-use, and efficiency, with the first two being particularly prominent.

Moreover, although the previous section discusses only the standard Newton iterative method for convenience, with a few brief modifications, our method can be iterated in other formats. For example, the simplified Newton method can be realized by moving the part of solving the Jacobian matrix (line 40 in Listing \ref{lst:code1}) outside the iteration's loop and, on top of that, adding an additional modified format about the Jacobian matrix, such as Broyden-rank-1 or Broyden-Fletcher-Shanme-rank-2, during the iteration corresponds to the quasi-Newton method. The above treatments degrade the convergence order but significantly reduce the computation cost for obtaining the Jacobian matrix via AD, which may have efficiency advantages in large-scale problems. More details can be found in our open-source code and will not be repeated here.

\section{Conclusions}
\label{sec:con}
In this paper, we have proposed a modified Newmark/Newton-Raphson method with automatic differentiation for nonlinear dynamics. This modified method significantly streamlines the acquisition of the precise Jacobian matrix during Newton iterations, effectively circumventing the cumbersome, error-prone, and time-consuming limitations of analytical differentiation while avoiding the numerical errors inherent in numerical differentiation. The proposed method provides a convenient and versatile tool for solving and analyzing complex nonlinear dynamic systems. The main conclusions are summarized as follows:

(1) The proposed method accurately computes the Jacobian matrix during Newton iterations through automatic differentiation, which inherently eliminates numerical errors introduced by numerical differentiation. Moreover, it effectively circumvents the tedious, time-consuming, and error-prone challenges of manual differentiation, as well as avoids expression swell issues inherent in symbolic differentiation, particularly when handling systems with complex nonlinearities.

(2) The implementation of the proposed method offers remarkable convenience and accessibility, requiring no additional processing of nonlinear terms. It only necessitates the provision of the linear mass matrix, linear damping matrix, linear stiffness matrix, external force matrix, and the expression of nonlinear terms from the nonlinear system's dynamic equations, thereby achieving excellent modularity.

(3) Simulation results on six nonlinear dynamic systems (the van der Pol system, the Duffing system, a nonlinear pendulum system, a 40 degrees of freedom system with complex nonlinear stiffness, a system with both complex nonlinear stiffness and damping, and a 284 degrees of freedom system with nonlinear stiffness) demonstrate that the proposed method accurately computes dynamic responses across different types of nonlinear systems, thereby validating its generality and accuracy.

There are many directions for future research. The current method solves ordinary differential equations , while future work could extend automatic differentiation  to handle differential algebraic equations that characterize multibody dynamics. Furthermore, as the present method provides numerical solutions that cannot capture unstable solutions in certain nonlinear dynamic systems, automatic differentiation could be incorporated into semi-analytical methods like the incremental harmonic balance (IHB) method and the harmonic balance-alternating frequency/time domain (HB-AFT) method \cite{chenNonlinearDynamicsAnalysis2023,changModifiedIHBMethod2023, chenGeneralEfficientHarmonic2024}to enable deeper analysis of unstable solutions and bifurcations in nonlinear systems.

\section*{Acknowledgements}
It is very grateful for National Natural Science Foundation of China (Grant Nos. U244120491, 12422213, 12372008), and the National Key R \& D Program of China (Grant No. 2023YFE0125900).

\section*{Code and data availability}
The proposed methodology of this study is openly available in GitHub at \url{https://github.com/shuizidesu/nnr-ad}.

\appendix
\section{Appendix}
\label{app:A}
$\textbf{M}_s=\frac{\mu Al}{840(1+\phi )^2} \left[
\begin{array}
{cccc}
m_1&m_2&m_3&m_4\\
&m_5&-m_4&m_6\\
& &m_1&-m_2\\
sym.& & &m_5\\
\end{array}
\right]
 + 
 \frac{\mu I_z}{30l(1+\phi )^2} \left[
      \begin{array}
      {cccc}
      m_7&m_8&-m_7&m_8\\
      &m_9&-m_8&m_{10}\\
      & &m_7&-m_8\\
      sym.& & &m_9\\
      \end{array}
      \right]$,

$\textbf{J}_s=\frac{\mu I_z}{15l(1+\phi )^2} \left[
      \begin{array}
      {cccc}
      m_7&m_8&-m_7&m_8\\
      &m_9&-m_8&m_{10}\\
      & &m_7&-m_8\\
      sym.& & &m_9\\
      \end{array}
      \right]$,
$\textbf{K}_s=\frac{{E}I_z}{l^3(1+\phi )} \left[
      \begin{array}
      {cccc}
      12&6l&-12&6l\\
      &(4+\phi)l^2&-6l&(2-\phi)l^2\\
      & &12&-6l\\
      sym.& & &(4+\phi)l^2\\
      \end{array}
      \right]$.

Where, $\phi=\frac{12{E}I_z}{\kappa Al^2}$, $m_1=312+588\phi+280\phi^2$, $m_2=(44+77\phi+35\phi^2)l$, $m_3=108+252\phi+140\phi^2$, $m_4=-(26+63\phi+35\phi^2)l$, $m_5=(8+14\phi+7\phi^2)l^2$, $m_6=-(6+14\phi+7\phi^2)l^2$, $m_7=36$, $m_8=(3-15\phi)l$, $m_9=(4+5\phi+10\phi^2)l^2$, $m_{10}=(-1+5\phi+5\phi^2)l^2$, $\kappa$ represents the shear correction factor of the shaft element, $\mu $ refers to the density of the shaft element,  $l$ denotes the length of the shaft element, $A$ represents the cross-sectional area of the shaft element, $E$ stands for the Young's modulus of the shaft element, and $I_z$ indicates the moment of inertia of the shaft element.

% To print the credit authorship contribution details
\printcredits

%% Loading bibliography style file
\bibliographystyle{model1-num-names}
% \bibliographystyle{cas-model2-names}

% Loading bibliography database
\bibliography{cas-paper-refs}

% Biography
% \bio{}
% Here goes the biography details.
% \endbio

% \bio{pic1}
% Here goes the biography details.
% \endbio

\end{document}